\documentclass[11pt,twoside,usenatbib]{article}

\usepackage{asp2010}

\resetcounters

\begin{document}

\title{Magnetic white dwarfs with debris disks}
\author{B. K\"ulebi,$^{1,2}$ 
        K.Y. Ek{\c s}i,$^{3}$
        P. Lor{\'e}n--Aguilar,$^{4,2}$
        J. Isern$^{1,2}$ and
        E. Garc{\'\i}a--Berro$^{4,2}$\\
\affil{$^1$Institut de Ci\`encies de l'Espai (CSIC), 
           Facultat de Ci\`encies, 
           Campus UAB, 
           Torre C5-parell, 
           08193 Bellaterra, 
           Spain}
\affil{$^{2}$Institute for Space Studies of Catalonia,
            c/Gran Capit\`a 2--4, Edif. Nexus 104,   
            08034  Barcelona, 
	    Spain}
\affil{$^{3}$Istanbul Technical University, 
	    Faculty of Science and Letters,
	    Physics Engineering Department,
	    Maslak 34469, Istanbul, Turkey}
\affil{$^{4}$Departament de F\'\i sica Aplicada, 
            Universitat Polit\`ecnica de Catalunya,  
            c/Esteve Terrades, 5,  
            08860 Castelldefels, Spain}}

\def\Rej{\mbox{RE{}\,J~0317-853}}
\newcommand{\Teff}{\hbox{$T_{\rm eff}$}}
\newcommand{\Msolar}{\mbox{\,$M_{\sun}$}}
\newcommand{\Rsolar}{\mbox{\,$R_{\sun}$}}
\newcommand{\Mch}{\mbox{\,$M_{\rm Ch}$}}
\newcommand{\M}{\mbox{M}}
\newcommand{\RWD}{\mbox{$R_{\rm WD}$}}
\newcommand{\MWD}{\mbox{$M_{\rm WD}$}}
\def\euve{\mbox{WD{}\,1658+441}}

\def\apj{ApJ}
\def\apjs{ApJS}
\def\apjl{ApJL}
\def\nat{Nature}
\def\aj{AJ}
\def\aap{A\&A}
\def\mnras{MNRAS}
\def\pasj{PASJ}
\def\nar{New Astron. Rev.}
\def\araa{ARAA}
\def\actaa{Acta Astron.}
\def\pasp{PASP}
\def\na{New Astron.}
\def\aapr{AAPR}

\begin{abstract}
It has long been accepted that a possible mechanism for explaining the
existence of  magnetic white  dwarfs is the  merger of a  binary white
dwarf system, as there are viable mechanisms for producing sustainable
magnetism  within the  merger  product.  However,  the  lack of  rapid
rotators  in  the magnetic  white  dwarf  population  has been  always
considered a problematic  issue of this scenario. In  order to explain
this discrepancy we build a model in which the interaction between the
magnetosphere of the star and the disk induces angular momentum transfer. Our
model predicts  that the magnetospheric interaction  of magnetic white
dwarfs with  their disks  results in a  significant spin down,  and we
show that the observed rotation  period of \Rej, which is suggested to
be a product of a double degenerate merger, can be reproduced.
\end{abstract}

\section{Introduction}
\label{sec:introduction}

Magnetic  white dwarfs (MWDs)  --- white  dwarfs with  field strengths
ranging from 1\,kG \citep{Jordanetal07} up  to 1\,GG \citep{Kawkaetal07,
Kulebietal09} ---  comprise more than $\sim$10\% of  all white dwarfs.
There  are two  possibilities  to account  for  the observed  magnetic
fields.  According to the  fossil field hypothesis, these white dwarfs
descend  from  Ap/Bp  stars  \citep{Angeletal81}, and  their  magnetic
fields are  remnants of  the previous evolution.   However, population
synthesis  studies   \citep{WickramasingheFerrario05}  show  that  the
observed number of  Ap/Bp stars is insufficient to  explain the number
of  MWDs, as  well as  other  properties of  their distribution.   One
mechanism which has been suggested as viable for producing
ultra-massive white dwarfs ($\MWD > 1.1 \Msolar$) is the merger of two
white dwarfs, which naturally explains  why MWDs are more massive than
their field counterparts \citep{Marshetal97, Kepleretal07}.

\citet{GarciaBerroetal12}  proposed  that  these magnetic  fields  are
generated  in   the  corona  above  the  merger   product  through  an
$\alpha\omega$  dynamo,  since  convection and  differential  rotation
exist  simultaneously.   Using  the  results  of  the  simulations  of
\citet{Loren-Aguilaretal09}  they showed  that very  powerful magnetic
fields can  be generated  in the hot  corona resulting from  a merger.
Also, to account for the  slow rotational velocities of the MWDs, they
suggested  magneto-dipole radiation  as a  possible solution  for this
drawback  of the  model.  However,  it is  likely  that magneto-dipole
radiation would not  be strong enough to slow down  the remnants of the
coalescences to the observed  rotation periods of MWDs.  Nevertheless,
if the  merger product  has high magnetic  dipole fields and  the disk
from the merger survives, it  is possible that the long-term evolution
of the system will involve  angular momentum exchange between the disk
and the star, depending on  the magnetic field strength of the central
object.  In this  work we  show  that this  angular momentum  exchange
successfully spins down the central MWD to the observed periods.

\section{Debris disk}

To model the  accretion disks which are remnants  of double degenerate
mergers,  we consider  a freely  expanding  thin disk  model which  is
similar  to  the  disks  formed   by  tidal  disruption  of  stars  by
super-massive black holes \citep{Cannizzoetal90} or supernova fallback
disks  around  young  neutron  stars, e.g.\  anomalous  X-ray  pulsars
\citep{Chatterjeeetal00,  Alpar01}.   In   this  type  of  model,  the
evolution of  the disk can be  described in terms of  its initial mass
and angular momentum \citep{Ertanetal09}.

The temporal evolution  of the surface mass density in the
disk is described by a diffusion equation \citep{Pringle81}:
\begin{equation}
\frac{\partial \Sigma }{\partial t}=\frac{3}{r}\frac{\partial }{\partial r}%
\left[ r^{1/2}\frac{\partial }{\partial r}\left( \nu \Sigma r^{1/2}\right) %
\right]   \label{diffuse}
\end{equation}
where $\Sigma$ is the surface mass density and $\nu $ is the turbulent
kinematic viscosity.   The rest of  the thin disk  structure equations
\citep{Franketal02} can be solved algebraically \citep{Cannizzoetal90}
to obtain a viscosity in  the form $\nu =Cr^{p}\Sigma ^{q}$ where $C$,
$p$ and  $q$ are  constants determined by  the opacity regime  --- see
also    \cite{Ertanetal09}.     For    this   type    of    viscosity,
Eq.~(\ref{diffuse})   has   the   self-similar  solutions   found   by
\citet{Pringle74}.   One  of  these  solutions conserves  the  angular
momentum of  the disk ($\dot{J}_{\mathrm{d}}=0$), and the  mass of the
disk evolves as a power-law  in time. In particular, for this solution
the mass flow rate at the inner boundary can be expressed as:
\begin{equation}
\dot{M}=\dot{M}_{0}\left( 1+\frac{t}{t_{0}}\right) ^{-\alpha }, \qquad 
\dot{M}_{0} = (\alpha-1)\frac{M_0}{t_0}
\label{Mdot}
\end{equation}
where $t_{0}$ is the  viscous timescale at some characteristic radius,
$M_0$ the  initial mass of the  disk and $\dot{M}_{0}$  is the initial
accretion rate. The  power-law index $\alpha$ can be  written in terms
of $p$  and $q$, which  in turn are  determined by the  opacity regime
\citep{Cannizzoetal90}.   For  electron  scattering  the  exponent  is
$\alpha=19/16$ and for the bound-free opacity regime is $\alpha=5/4$.

The most important  parameter governing the spin evolution  of MWDs is
the viscous timescale  $t_0$, which is determined by  the initial mass
and  angular  momentum  of  the  disk  \citep{Ertanetal09}.   For  our
calculations we  adopt a metal  disk in the bound-free  opacity regime
with $\alpha_{\rm SS}=0.1$, for which the viscous timescale is
\begin{equation}
t_0 \simeq 15\,{\rm s}\,\left(\frac{j_0}{10^{18}\,{\rm cm}^2 \, {\rm s}^{-1}}
\right)^{25/7}\left(\frac{M_0}{10^{-1}\,\Msolar}\right)^{-3/7}
\end{equation}
where $j_0=J_0/M_0$  is the specific  angular momentum. Note  that the
viscous timescale depends strongly on $j_0$.

The  way in  which the  star interacts  with the  disk depends  on the
interplay between the angular velocity  of the star $\Omega_{\rm WD }$
and  the  keplerian   velocity  at  the  inner  radius   of  the  disk
$\Omega_{\rm  K}$. This  is  parametrized by  the fastness  parameter
$\omega_{\ast  }=\Omega_{\rm  WD}  /\Omega_{\rm  K}(R_{\rm in})$, from  which  the
torque transfer can be computed:
\begin{equation}
N_{\rm d} = n(\omega_{\ast})N_0.
\end{equation}
where $N_0=\Omega_{\rm  K}\left(R_{\rm in}\right) R_{\rm in}^2\dot{M}$
and   $n(\omega_{\ast})$   is    called   the   dimensionless   torque
\citep{GhoshLamb79,Lamb88,Lamb89}.   Following   \citet{Alpar01}   and
\citet{Eksietal05}  we  employ  $n=1-\omega  _{\ast}$.  If  the  inner
radius of the disk goes beyond the light cylinder radius, the disk can
no longer torque  the star via the magnetosphere,  $N_{\rm d} =0$.  In
this   so-called   ejector  stage   the  star   will   spin-down  via
magneto-dipole radiation

\section{Results}
\label{sec:results}

To compare  our model with observations,  we first conducted  a set of
simulations in which the adopted masses of the central remnant and the
disk,  the angular  velocity of  the uniformly  rotating star  and the
angular  momentum of the  disk were  those of  the SPH  simulations of
\citet{Loren-Aguilaretal09}.   For a  given simulation,  we calculated
the spin evolution of the MWD for different polar field strengths.  To
determine the total integration time we used the simple cooling law of
\citet{Mestel65}  and  estimated  the  cooling age  at  the  otherwise
typical observed value of \Teff$\sim40\,000$~K.  The exact value of of
the  initial core  temperature  depends  on the  mass  of the  merging
components.  However, the exact  value of the initial core temperature
has virtually  no consequences for  the age estimate, as  white dwarfs
cool very rapidly  during the early phases of  evolution.  Under these
conditions   the  cooling   ages  for   MWDs  of   masses   0.8,  1.2,
1.4\,\Msolar\ MWDs were 13, 50 and 100~Myr respectively.

The evolution of the angular velocity of the star is determined by the
fastness  parameter, $\omega_\ast$.  Initially  the accretion  rate is
very large.  Hence,  the magnetospheric radius is very  small, and the
inner radius of the disk is equal to the stellar radius. In this case,
the fastness parameter is smaller  than 1. Thus, the  remnants of mergers  are fast rotators,
and are rotating at almost  one third of the critical angular velocity
of  the  star  $\omega_\ast\sim0.33$~s$^{-1}$.   As  time  passes  by,
angular  momentum  is  transferred  onto  the star  and  the  fastness
parameter approaches 1. If the mass of the white dwarf is not close to
\Mch\ the star does not reach critical rotation and this spin-up stage
continues till  $\omega_\ast> 1$.  At  this point the  magnetic torque
from the  disk starts to  spin down the  star.  As the accretion rate 
decreases, the magnetospheric radius and hence  $\omega_{\ast}$, 
increase rapidly. This results in a stronger torque on the star which 
eventually forces it to reach torque equilibrium at $\omega_{\ast} \simeq 1$. 
This phase is called the tracking phase \citep{Chatterjeeetal00}. 
This is a quasi-equilibrium phase because the mass transfer rate keeps
declining and the equilibrium period, at which $\omega_{\ast} \simeq 1$, increases. Hence, accretion continues while the
star spins down. This second  phase of  torque  transfer is
observed in all our simulations and the strength of the magnetic field
determines when  this stage  is reached. If  the polar  magnetic field
$B_p$  is strong,  the Alfv\'en  radius and  $\omega_{\ast}$  are also
larger.  In  all our  simulations there is  a limiting  magnetic field
strength which determines  whether an MWD will enter  this strong spin
down phase or not, given the  same cooling age.  If the magnetic field
is close  to this limiting value,  $B_{\rm lim}$, the  final period of
the star is very sensitive to the exact value of $B_p$. In particular,
for  a field  strength  of $B_{\rm  lim}$,  a 10\%  difference in  the
initial field strength results in a change by a factor $\sim 2$ of the
final rotation period.  Finally, after this strong spin down phase, if
the magnetic field strengths and  cooling age are adequate, the system
reaches quasi-equilibrium around $\omega_\ast\la 1$.

The period  reached after the  tracking phase depends on  time, dipole
field strength, angular  momentum of the disk, mass  and radius of the
star, but not on the mass  of the disk. Building upon this, and taking
into  account   that  there   are  considerable  unknowns   about  the
super-Eddington accretion  phase, we  also built models  with low-mass
disks.   This approximation  is  acceptable since  this  phase of  the
evolution is  short ($10^4$~yr) with respect to  the evolutionary time
of the observed population of MWDs ($10^7-10^8$~yr).  For these models
we assumed a disk mass of  $10^{-3}\,\Msolar$ --- the mass of the disk
when the  Eddington accretion phase  finishes and accretion  driven by
the evolution of angular momentum takes over --- while the rest of the
material was incorporated onto  the remnant white dwarf.  We performed
this test for  a 1.2\,\Msolar\ MWD, and the  results were identical to
those of obtained in the previous set of simulations, for cooling ages
longer than  $\sim10^7\,$yr and $B_p > B_{\rm  lim}\sim 10\,$MG, which
apply for most high-field MWDs with known periods.

\begin{figure}
\includegraphics[width=0.9\columnwidth]{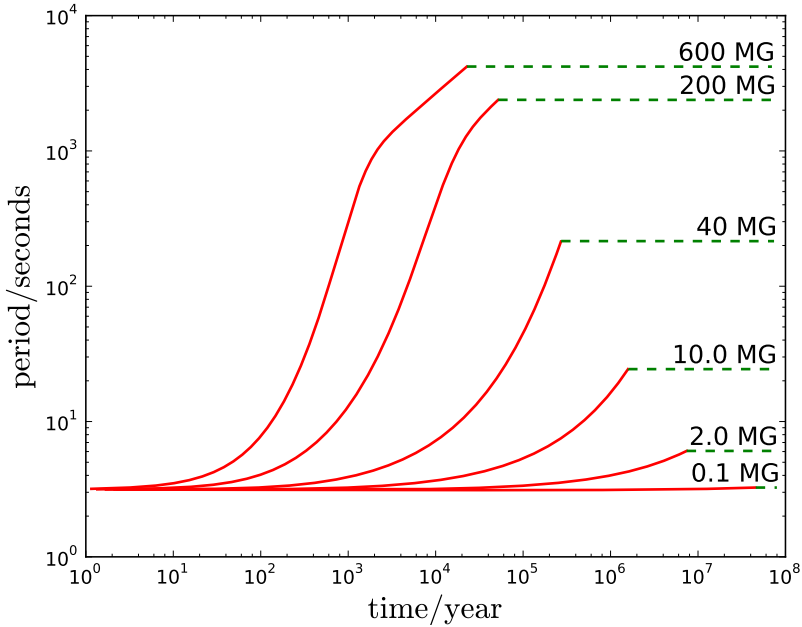}
\centering
  \caption{The evolution of the period  for a 1.30\,\Msolar\ MWD for a
    turbulence turn-off  temperature $T_p=1000\,$K.  The  dashed lines
    show where  the torque transfer from the  disk becomes ineffective
    and magneto-dipole spin down takes over.}
  \label{fig:Pvst02}
\end{figure}

The lifetime  of a MWD is expected  to be long when  compared with the
typical cooling timescale  of the disk. Thus, it  is natural to wonder
whether or  not turbulence  in the disk  can be sustained  during long
periods of time.  The cooling timescale of the  disk depends primarily
on its degree of  the ionization.  \citet{Menouetal01} showed that for
thin disks made of C, thermal equilibrium is reached near $5\,000\,$K,
and  that  for  all  compositions recombination  always  occurs  below
$\sim1\,000\,$K.  Although this temperature  is high, it is known that
these disks can be partially ionized at even lower temperatures due to
several  other  processes, like  cosmic  ray  heating, collisions  and
charged-particle  mobility.  \citet{InutsukaSano05}  carefully studied
all these effects and reached the conclusion that at temperatures $\la
1\,000\,$K,   magneto-rotational   instability  \citep{BalbusHawley91}
could be still effective.

To implement the effects of the turbulent life time, we calculated the
temperature at  the inner radius  of the disk  at each time  step.  If
this temperature was  smaller than $T_p = 1\,000\,$K,  we assumed that
the disk was completely inactive and accretion was switched-off. Thus,
the MWD  immediately enters in  the so-called ejector phase  and spins
down only by magneto-dipole  radiation.  To compare these calculations
to   the    case   of   \Rej\    ---   with   a   period    of   721~s
\citep{Barstowetal95,Ferrarioetal97}              ---              and
\euve\ \citep{Schmidtetal92}  we applied  this procedure to  the model
with $M=1.30\,\Msolar$.   The evolution of  the periods for  MWDs with
different dipole fields is  shown in Fig.\,\ref{fig:Pvst02}.  We found
that  although   the  magneto-rotational  instability   is  long-lived
($\sim10^5-10^7\,$years),  the disk  becomes inactive  at  the current
ages  of  these  objects  ($\sim10^8\,$years).   Moreover,  since  the
magneto-dipole torque is not as strong as the magnetic torque from the
disk,  the  final  period is  determined  by  the  time at  which  the
magneto-rotational  instability is  turned off,  which depends  on the
accretion rate.

\section{Conclusions}

For  MWDs with  field strengths  between 50\,MG  and 1\,GG,  and $T_p$
ranging from  $\sim300\,$K to 1\,000~K, the final  periods reached are
around $10^3-10^4$\,s (see  Fig.\,\ref{fig:Pvst02}), in agreement with
the period of \Rej. However this  is not true for the MWD \euve\ which
has a  field strength  of $\approx2\,$MG and  undetected ---  hence, a
long  ($> 10^5\,$s)  --- period  \citep{Shtoletal97}.  \Rej\  has been
suggested to  be a  merger remnant  due to its  mass, cooling  age and
strong magnetism.
Our simple evolutionary models of the coupling between the white dwarf 
and the surrounding disk show that the disk may remove angular momentum 
from the central white dwarf for reasonable choices of the free parameters, 
and thus these models are able to explain the observed properties of REJ 0317-853.

\acknowledgements  This  research   was  supported  by  MICINN  grants
AYA2011--23102  and  AYA08-1839/ESP,  by  the  ESF  EUROCORES  Program
EuroGENESIS (MI\-CINN  grant EUI2009-04170), by  grants 2009SGR315 and
2009SGR1002 of the Generalitat de  Catalunya and by the European Union
FEDER funds.

\bibliographystyle{asp2010}
\bibliography{./kulebi}

\begin{thebibliography}{}
\expandafter\ifx\csname natexlab\endcsname\relax\def\natexlab#1{#1}\fi
\expandafter\ifx\csname url\endcsname\relax
  \def\url#1{\texttt{#1}}\fi
\expandafter\ifx\csname urlprefix\endcsname\relax\def\urlprefix{URL }\fi
\providecommand{\eprint}[2][]{\url{#2}}

\bibitem[{{Alpar}(2001)}]{Alpar01}
{Alpar}, M.~A. 2001, \apj, 554, 1245

\bibitem[{{Angel} et~al.(1981){Angel}, {Borra}, \& {Landstreet}}]{Angeletal81}
{Angel}, J.~R.~P., {Borra}, E.~F., \& {Landstreet}, J.~D. 1981, \apjs, 45, 457

\bibitem[{{Balbus} \& {Hawley}(1991)}]{BalbusHawley91}
{Balbus}, S.~A., \& {Hawley}, J.~F. 1991, \apj, 376, 214

\bibitem[{{Barstow} et~al.(1995){Barstow}, {Jordan}, {O'Donoghue}, {Burleigh},
  {Napiwotzki}, \& {Harrop-Allin}}]{Barstowetal95}
{Barstow}, M.~A., {Jordan}, S., {O'Donoghue}, D., {Burleigh}, M.~R.,
  {Napiwotzki}, R., \& {Harrop-Allin}, M.~K. 1995, \mnras, 277, 971

\bibitem[{{Cannizzo} et~al.(1990){Cannizzo}, {Lee}, \&
  {Goodman}}]{Cannizzoetal90}
{Cannizzo}, J.~K., {Lee}, H.~M., \& {Goodman}, J. 1990, \apj, 351, 38

\bibitem[{{Chatterjee} et~al.(2000){Chatterjee}, {Hernquist}, \&
  {Narayan}}]{Chatterjeeetal00}
{Chatterjee}, P., {Hernquist}, L., \& {Narayan}, R. 2000, \apj, 534, 373

\bibitem[{{Ek{\c s}i} et~al.(2005){Ek{\c s}i}, {Hernquist}, \&
  {Narayan}}]{Eksietal05}
{Ek{\c s}i}, K.~Y., {Hernquist}, L., \& {Narayan}, R. 2005, \apjl, 623, L41

\bibitem[{{Ertan} et~al.(2009){Ertan}, {Ek{\c s}i}, {Erkut}, \&
  {Alpar}}]{Ertanetal09}
{Ertan}, {\"U}., {Ek{\c s}i}, K.~Y., {Erkut}, M.~H., \& {Alpar}, M.~A. 2009,
  \apj, 702, 1309

\bibitem[{{Ferrario} et~al.(1997){Ferrario}, {Vennes}, {Wickramasinghe},
  {Bailey}, \& {Christian}}]{Ferrarioetal97}
{Ferrario}, L., {Vennes}, S., {Wickramasinghe}, D.~T., {Bailey}, J.~A., \&
  {Christian}, D.~J. 1997, \mnras, 292, 205

\bibitem[{{Frank} et~al.(2002){Frank}, {King}, \& {Raine}}]{Franketal02}
{Frank}, J., {King}, A., \& {Raine}, D.~J. 2002, {Accretion Power in
  Astrophysics: Third Edition} (Cambridge University Press)

\bibitem[{{Garc{\'{\i}}a-Berro} et~al.(2012){Garc{\'{\i}}a-Berro},
  {Lor{\'e}n-Aguilar}, {Aznar-Sigu{\'a}n}, {Torres}, {Camacho}, {Althaus},
  {C{\'o}rsico}, {K{\"u}lebi}, \& {Isern}}]{GarciaBerroetal12}
{Garc{\'{\i}}a-Berro}, E., {Lor{\'e}n-Aguilar}, P., {Aznar-Sigu{\'a}n}, G.,
  {Torres}, S., {Camacho}, J., {Althaus}, L.~G., {C{\'o}rsico}, A.~H.,
  {K{\"u}lebi}, B., \& {Isern}, J. 2012, \apj, 749, 25

\bibitem[{{Ghosh} \& {Lamb}(1979)}]{GhoshLamb79}
{Ghosh}, P., \& {Lamb}, F.~K. 1979, \apj, 234, 296

\bibitem[{{Inutsuka} \& {Sano}(2005)}]{InutsukaSano05}
{Inutsuka}, S.-i., \& {Sano}, T. 2005, \apjl, 628, L155

\bibitem[{{Jordan} et~al.(2007){Jordan}, {Aznar Cuadrado}, {Napiwotzki},
  {Schmid}, \& {Solanki}}]{Jordanetal07}
{Jordan}, S., {Aznar Cuadrado}, R., {Napiwotzki}, R., {Schmid}, H.~M., \&
  {Solanki}, S.~K. 2007, \aap, 462, 1097

\bibitem[{{Kawka} et~al.(2007){Kawka}, {Vennes}, {Schmidt}, {Wickramasinghe},
  \& {Koch}}]{Kawkaetal07}
{Kawka}, A., {Vennes}, S., {Schmidt}, G.~D., {Wickramasinghe}, D.~T., \&
  {Koch}, R. 2007, \apj, 654, 499

\bibitem[{{Kepler} et~al.(2007){Kepler}, {Kleinman}, {Nitta}, {Koester},
  {Castanheira}, {Giovannini}, {Costa}, \& {Althaus}}]{Kepleretal07}
{Kepler}, S.~O., {Kleinman}, S.~J., {Nitta}, A., {Koester}, D., {Castanheira},
  B.~G., {Giovannini}, O., {Costa}, A.~F.~M., \& {Althaus}, L. 2007, \mnras,
  375, 1315

\bibitem[{{K{\"u}lebi} et~al.(2009){K{\"u}lebi}, {Jordan}, {Euchner},
  {G{\"a}nsicke}, \& {Hirsch}}]{Kulebietal09}
{K{\"u}lebi}, B., {Jordan}, S., {Euchner}, F., {G{\"a}nsicke}, B.~T., \&
  {Hirsch}, H. 2009, \aap, 506, 1341

\bibitem[{{Lamb}(1988)}]{Lamb88}
{Lamb}, D.~Q. 1988, in Polarized Radiation of Circumstellar Origin, edited by
  G.~V. {Coyne}, A.~M. {Magalhaes}, A.~F. {Moffat}, R.~E. {Schulte-Ladbeck}, \&
  S.~{Tapia}, 151

\bibitem[{{Lamb}(1989)}]{Lamb89}
{Lamb}, F.~K. 1989, in Timing Neutron Stars, edited by {H.~{\"O}gelman \&
  E.~P.~J.~van den Heuvel} (Vatican Obs.), 649

\bibitem[{{Lor{\'e}n-Aguilar} et~al.(2009){Lor{\'e}n-Aguilar}, {Isern}, \&
  {Garc{\'{\i}}a-Berro}}]{Loren-Aguilaretal09}
{Lor{\'e}n-Aguilar}, P., {Isern}, J., \& {Garc{\'{\i}}a-Berro}, E. 2009, \aap,
  500, 1193

\bibitem[{{Marsh} et~al.(1997){Marsh}, {Barstow}, {Buckley}, {Burleigh},
  {Holberg}, {Koester}, {O'Donoghue}, {Penny}, \& {Sansom}}]{Marshetal97}
{Marsh}, M.~C., {Barstow}, M.~A., {Buckley}, D.~A., {Burleigh}, M.~R.,
  {Holberg}, J.~B., {Koester}, D., {O'Donoghue}, D., {Penny}, A.~J., \&
  {Sansom}, A.~E. 1997, \mnras, 287, 705

\bibitem[{{Menou} et~al.(2001){Menou}, {Perna}, \& {Hernquist}}]{Menouetal01}
{Menou}, K., {Perna}, R., \& {Hernquist}, L. 2001, \apj, 559, 1032

\bibitem[{{Mestel}(1965)}]{Mestel65}
{Mestel}, L. 1965, {Stars in Stellar Systems} (University of Chicago
  Press,~Chicago, 1965)

\bibitem[{{Pringle}(1974)}]{Pringle74}
{Pringle}, J.~E. 1974, Ph.D. thesis, , Univ.~Cambridge, (1974)

\bibitem[{{Pringle}(1981)}]{Pringle81}
--- 1981, \araa, 19, 137

\bibitem[{{Schmidt} et~al.(1992){Schmidt}, {Bergeron}, {Liebert}, \&
  {Saffer}}]{Schmidtetal92}
{Schmidt}, G.~D., {Bergeron}, P., {Liebert}, J., \& {Saffer}, R.~A. 1992, \apj,
  394, 603

\bibitem[{{Shtol'} et~al.(1997){Shtol'}, {Valyavin}, {Fabrika}, {Bychkov}, \&
  {Stolyarov}}]{Shtoletal97}
{Shtol'}, V.~G., {Valyavin}, G.~G., {Fabrika}, S.~N., {Bychkov}, V.~D., \&
  {Stolyarov}, V.~A. 1997, Astronomy Letters, 23, 48

\bibitem[{{Wickramasinghe} \& {Ferrario}(2005)}]{WickramasingheFerrario05}
{Wickramasinghe}, D.~T., \& {Ferrario}, L. 2005, \mnras, 356, 1576

\end{thebibliography}

\end{document}